# Multimodal Affect Recognition using Kinect

AMOL S. PATWARDHAN, Louisiana State University
GERALD M. KNAPP, Louisiana State University

Affect (emotion) recognition has gained significant attention from researchers in the past decade. Emotion-aware computer systems and devices have many applications ranging from interactive robots, intelligent online tutor to emotion based navigation assistant. In this research data from multiple modalities such as face, head, hand, body and speech was utilized for affect recognition. The research used color and depth sensing device such as Kinect for facial feature extraction and tracking human body joints. Temporal features across multiple frames were used for affect recognition. Event driven decision level fusion was used to combine the results from each individual modality using majority voting to recognize the emotions. The study also implemented affect recognition by matching the features to the rule based emotion templates per modality. Experiments showed that multimodal affect recognition rates using combination of emotion templates and supervised learning were better compared to recognition rates based on supervised learning alone. Recognition rates obtained using temporal feature were higher compared to recognition rates obtained using position based features only.

Categories and Subject Descriptors: **I.2.10 [Artificial Intelligence]**: Vision and Scene Understanding; **I.4.8 [Image Processing and Computer Vision]**: Scene Analysis; **I.5.4 [Pattern Recognition]**: Applications

## 1. INTRODUCTION

Multimodal techniques have recently become a focus in affect (emotion) recognition research. Modality is defined as the channel of input data. Data from facial expressions, head position, hand gesture, body posture, sound and physiological response such as heart beat, pulse, sweat, body temperature can be used as a modality. A survey on the state of the art in affect recognition using hand, arm, torso and body movement has been provided in (Stathopoulou & Tsihrintzis, 2011) indicating research focus shifting towards the area of multimodal affect recognition instead of affect detection using only facial expression analysis. Researchers have focused on unimodal or bimodal affect recognition using features from audio and visual modalities. A recent survey (Calvo & D'Mello, 2010) reviewed various studies focusing on basic emotions, facial expressions, feelings, social interaction, body language and neural circuitry. The survey also analyzed existing affect detection systems and emerging research on multimodal affect recognition. Modeling emotions, challenges in multimodal emotion detection, fusion methods and emerging applications of affective computing has been discussed in a research (Jaimes & Sebe, 2007). It has been shown that data from more than one modality provides better classification rates compared to individual modality. Researchers (Gunes & Piccardi, 2005) have used a multi frame post integration approach in affect analysis of facial and body display and showed improved recognition accuracy compared to only using face as modality. Researchers (Tan & Nareyek, 2009) showed integrating various modalities for affect recognition, provided better accuracy over individual modalities and also identified that hand gestures aided in emotion recognition from posture. A study (Li & Jarvis, 2009) proposed a multimodal gesture recognition and body pose estimation system from the viewpoint of a robot. The study used a range and stereo camera for depth data and sensing pose and gesture respectively. Research done in (Amelynck, Grachten, van Noorden, & Leman, 2011) used inertial sensors to detect arm gestures and perform music retrieval based on the recognized affect. Researchers have used geometric features and supervised learning techniques such as Support Vector Machines (SVM) for recognition of emotions. Grid based Geometric deformation features and classification using SVM was proposed in (Kotsia & Pitas, 2007). In a study (Suprijanto, Sari, Nadhira, Merthayasa, & Farida, 2009) researchers have used facial expressions and biometric signals for emotion recognition. The study discusses shape, intensity, hybrid models and features based on geometric properties of the eyes, light intensity and appearance of the eye. To the best of our knowledge no study has focused on using emotion templates to augment supervised learning results. In a previous work that used such rule based templates (Patwardhan & Knapp, 2013) only location based features and limited set of rules were incorporated. It also performed analysis using happiness as the candidate emotion for the pilot study.

The key objectives of this research are to 1) compare the affect recognition rates obtained from supervised learning alone with the recognition rates obtained by combining supervised learning and rule based emotion templates; 2) study the recognition rates by using not just position based features but also temporal features across frames containing joint and feature tracking data; and 3) extend the multimodal affect recognition prototype to recognize multiple emotions based on a confidence level threshold instead of attributing the user interaction to only a single emotion.

In this research audio and visual modalities were analyzed; physiological input was not considered. The Microsoft Kinect was utilized for feature tracking. The Kinect provides both RGB color and depth data, is readily available and cost effective, and easy to configure and setup. Additionally, it has well supported and robust software development kit (SDK) from Microsoft. Kinect uses infrared technology and hence eliminates the need to attach any sensors to the interacting user. The device can be connected directly to a computer without having to purchase the game console. In this research, the skeletal data and the face recognition capability in Kinect was used. Skeletal data is valuable in tracking location and movement of joints in human body. The Kinect SDK also contains a face recognition application programming interface (API) which provides the ability to track over 100 key facial feature points. Additionally, the Kinect provides speech recognition capability. In this research the body posture, hand movement and facial expression data were analyzed to match with a rule based emotion template to recognize the subject's emotion. Happiness, surprise, fear, anger, sadness and disgust were considered as the candidate emotions.

This research proposes a new set of rules and temporal features across 10 frames to track movement of each feature point. A rule based template was created for each combination of emotion and modality instead of a template detecting only happiness from face modality. Additionally, past studies have primarily focused on attributing only one emotion as the classification output. In this research we detect multiple emotions having confidence level above an experimentally determined threshold, as a person may exhibit multiple concurrent emotions (surprise and happiness at the same time, for example). Identifying multiple emotions helps describe the emotional state of an individual more accurately and can be used by a computer system to adjust the response effectively, making it more functionally appealing to the user.

## 2. FEATURE EXTRACTION
### 2.1 Location based features

This research used a feature vector composed of the x, y coordinates of tracked points, the angle made by pairs of tracked points with the horizontal axis, and the Euclidean distance between pairs of tracked points. This feature vector was based on tracked points obtained from only one frame in a moment of time. Feature vector definition and feature extraction technique for hand, head, shoulder, body and face has been provided in (Gunes 2005) and (Patwardhan & Knapp, 2013). The feature vector for hand modality is defined as follows:

$$FV_{ha} = \{P_{1,x}(n), P_{1,y}(n), \ldots P_{8,y}(n), d(P_1(n), P_2(n)), \ldots, d(P_7(n), P_8(n))$$
$$, \theta(P_1(n), P_2(n)), \ldots, \theta(P_7(n), P_8(n))\} \qquad (1)$$

The value $P_{i,c}(n)$ is the x, y co-ordinate value of $i^{th}$ tracked point in the $n^{th}$ frame where $c \epsilon \{x, y\}$; d is the Euclidean distance between two distinct tracked points. $\theta$ is the angle between two distinct tracked points. The feature vector is based on purely position of tracked points and meta-data about change in location of tracked point is not available in the feature vector. To overcome this drawback, temporal feature vectors were developed as described in the next section.

### 2.2 Temporal features

This section defines the temporal feature vectors used by the emotion template based recognition, followed by feature definition for SVM based recognition module. The motivations behind using temporal features instead of positional features are twofold. First, the location based features depend on tracked points only in a given point of time. The tracked points are obtained from the skeletal frame data generated from the Kinect. Second, even if these features are sufficient to identify head movement, hand gestures, body postures or changes in facial expressions, the feature vector can be made more robust by introducing temporal elements. Last, it is important to capture the movement of each tracked point across multiple frames in order to completely define and identify a gesture or facial expression. The temporal data can then be used in each emotion template to further improve emotion prediction accuracy for a small computational overhead. As a result, in addition to location based features, the study also proposed extracting temporal features across multiple frames and adding rules to each emotion template to infer emotions based on movement of tracked points. A window of 10 frames was used for feature extraction. The movement of each feature defined in the location based feature section was tracked across the 10 frames. The temporal feature vector obtained from hand as the modality is defined as follows:

*$FVTn = \{FVT_{ha1}, \ldots, FVT_{han}\}$ denotes the set of n hand feature vectors used as input by the emotion matching algorithm. This set is analyzed to track the movement of joints across multiple frames and evaluating rules that define whether the movement of the tracked joints represents a gesture associated with an emotion.*

$$FVT_{ha} = \{PT_{1,x}(n), PT_{1,y}(n), \ldots PT_{6,y}(n)\} \qquad (2)$$

The temporal feature vector for hand modality $FVT_{ha}$ is the set of x and y co-ordinates of tracked joints. The six tracked joints for hand movement are left shoulder, right shoulder, left elbow, right elbow, left wrist and right wrist. The feature vectors for the other modalities such as head movement, facial expressions, body posture were defined using the same technique, except different tracked points. For head modality, movement of top of skull, right and left cheeks and bottom of chin were tracked. For the facial expressions, left and right cheeks, left, right and center of both lips, left, right and center of both eyes, left, right and center of both eyebrows were tracked. For the body modality, center of shoulder, spine, left, right and center of hip in addition to all the points from hand modality were tracked. In case of SVM based recognition the movement of tracked points across consecutive frames was measured and used as feature vector. The two important features obtained from the temporal data across frames were angle and magnitude of velocity of the tracked points. The feature calculation is described in equation as follows:

$$FVTSVM_{ha} = \{PT_1(\theta), PT_1(v), \ldots PT_6(v)\} \quad (3)$$

$FVTSVM_{ha}$ is the feature vector defining the set of features used for tracking hand movement across frames. This feature vector is used for training and detection in SVM based classification. The feature vector is a set of velocity vectors represented by magnitude and orientation for each tracked point. $PT_n(\theta)$ is the direction component which is the angle made by the $n^{th}$ tracked point in the current frame and the previous frame with the horizontal and is calculated using the x and y co-ordinates of the tracked point. $PT_n(v)$ is the magnitude component of the velocity vector calculated using displacement of y co-ordinates relative to displacement in x co-ordinate of the $n^{th}$ tracked point. Thus the feature vector defines the movement of various tracked joints and facial points obtained from Kinect in terms of magnitude and orientation.

## 3. EMOTION RECOGNITION

A classifier for each modality was trained using support vector machine. Location based and temporal features discussed in the preceding section were used for training. In addition to supervised learning technique, this research used emotion detection based on matching the features with a rule based emotion template for each combination of emotion and modality. In a previous work, a pilot study was performed using an emotion template for recognition of happiness, which proved that emotion template based detection was feasible. The current research extends the concept to other emotions and modalities. This research also used temporal features in addition to location based features as inputs to the template and studied the resulting effect on the classification rates. Additional rules were added to the emotion template to study the influence on the classification rates obtained by combining the results with the results obtained using the supervised learning technique.

### 3.1 Emotion Template Matching

An emotion template is a set of rules that define a particular emotion based on feature data available from a modality. The matching process evaluates each rule in the emotion template and if the tracked data satisfies all the rules then the output of the prediction is the emotion defined by the template. The emotion matching module consists of two stages. The first stage analyzes the tracked feature data on a per frame basis. The data from each frame is stored in a buffer which is passed as input to the emotion matching module. Each index location in the buffer represents tracked data from a frame. The data from the buffer is read and evaluated against the emotion templates inside a loop. This stage evaluates the rules in the templates that evaluate the position based features per frame. The outcome of each evaluation is a Boolean indicating whether the emotion defined by the template was detected or not. The second stage of the template matching process analyzes the tracked feature data across multiple frames by taking into account the entire buffer content. This is achieved by passing the entire buffer and not just content at a specific index location in the buffer to a temporal rule evaluator. For instance, the temporal rule evaluator examines the movement of a tracked left wrist across multiple frames stored in the buffer. An emotion template contains several calls to individual rule evaluators. The outcome of an emotion template matching process for a specific emotion is true only if all the positional rules and temporal rules that define the emotion are satisfied. The next section describes the positional rules and temporal rules in detail.

### 3.2 Positional Rules

According to a rule from the emotion template defining happiness, if the person raised the arms above the head then the matching algorithm identifies it as an expression of joy. According to a rule from the emotion template defining sadness, if the person raises the arms above shoulder and holds it behind the head the matching algorithm identifies it as an expression of disappointment and associates it to the emotion of sadness. The equations below describe four rules which are part of a larger set of rules defining an emotion template for happiness.

$$Rule\ 1_{happy} = \{PT_{re,y}(n) > PT_{rs,y}(n)\} \tag{4}$$

$$Rule\ 2_{happy} = \{PT_{le,y}(n) > PT_{ls,y}(n)\} \tag{5}$$

$$Rule\ 3_{happy} = \{PT_{lw,y}(n) > PT_{le,y}(n)\} \tag{6}$$

$$Rule\ 4_{happy} = \{PT_{rw,y}(n) > PT_{re,y}(n)\} \tag{7}$$

$PT_{rs,y}$, $PT_{re,y}$, $PT_{rw,y}$, $PT_{ls,y}$, $PT_{le,y}$, $PT_{lw,y}$ are the y co-ordinates of the tracked right shoulder, right elbow, right wrist, left shoulder, left elbow and left wrist respectively. Rule $N_e$ define an emotion template where N is the number of rules and e is the emotion class. Rule1 to Rule 4 are useful to detect if the subject has raised the arms above shoulder in joy. These rules are used to evaluate only position based data such as co-ordinate values from the tracked points per frame.

### 3.3 Temporal Rules

A rule evaluating temporal feature defines simultaneous upward and downward motion of elbow and wrist joints above the shoulder as an expression of happiness. Similarly, a rule defining sadness evaluates movement of head shaking left or right indicating disapproval. The same rule can be part of multiple templates since shaking head left or right can represent expression of frustration and anger.

$$Temporal\ Rule\ 1_{happy} = |PT_{re,y}(n)| > cy; |PT_{re,y}(n) > PT_{rs,y}(n) + ty \tag{8}$$

$$Temporal\ Rule\ 2_{happy} = |PT_{rw,x}(n)| > cx;\ s.t.PT_{rw,x}(n) > PT_{re,x}(n) + tx \tag{9}$$

$PT_{rs,y}$, $PT_{re,y}$ are the y coordinates of right shoulder and right elbow. $PT_{rw,x}$, $PT_{re,x}$ are the x coordinates of right wrist and right elbow. Temporal Rule 1 counts the number of times right elbow moves above the right shoulder by a distance ty and checks if the count is greater than cy. It defines how many times the right elbow was raised above right shoulder and by how much. If the tracked points are within the defined threshold, it means the Temporal Rule 1 evaluated to true. Similarly, Temporal Rule 2 counts the number of times right wrist moved across right elbow towards right by a distance tx and checks if the count is greater than cx. It defines how many times the wrist was moved back and forth and by how much.

## 4. AFFECT RECOGNITION SYSTEM

### 4.1 ARUMM (Affect recognition using multiple modalities)

An affect recognition system was developed using Windows Presentation Foundation (WPF) and Microsoft Kinect SDK. The windows based program allowed capturing skeletal data and facial expressions, extract features and feed it to the rule based emotion template matching component for affect recognition. This version was an improvement to a previously developed system prototype. The system was named ARUMM (Affect recognition using multiple modalities).

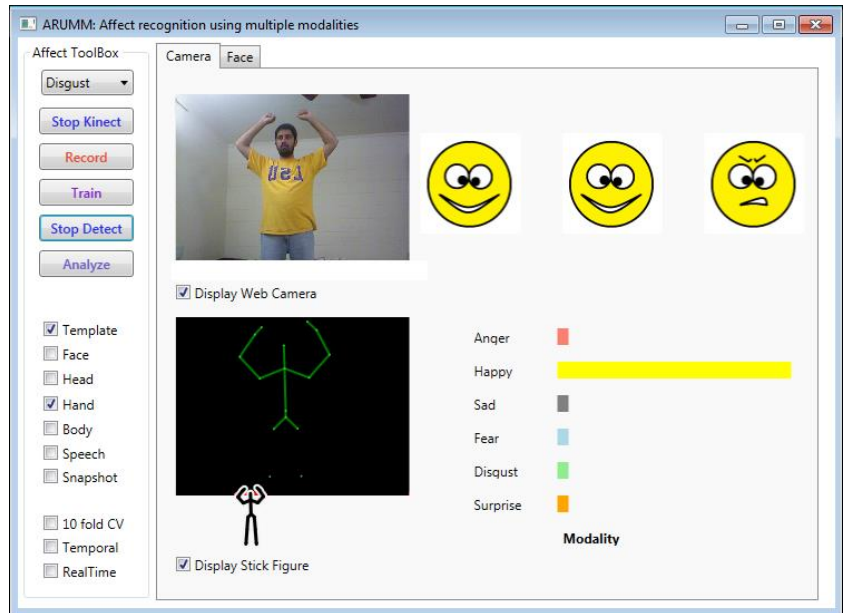

Fig. 1. ARUMM version 1.0.

Figure 1 shows a screen shot of the ARUMM version 1.0 system. The dashboard displays the camera output, the skeletal output and the top three emotions that have an empirical probability greater than the confidence level threshold, detected at runtime by the affect recognition component. The emotional state is represented using emoticons and a horizontal bar graph showing the confidence level of the detected emotion. The example above shows happiness as the emotion having the highest confidence level. Anger is detected as the emotion with the next highest confidence level above threshold.

### 4.2 Fusion Method

The fusion strategy employed in this research was late binding of outputs from multiple modalities. Emotion outcome from each available modality was combined at the decision level. In addition to only one emotion outcome from multiple inputs, this study proposes calculating emotions based on confidence level. This approach extends the multimodal affect recognition capability to further identify more than one emotion. An event driven fusion approach was adopted in this research. This approach was useful for multimodal affect recognition. Additionally, this research combined the SVM technique with the emotion template based recognition. For the purpose of determining the final emotion, the confidence level of outcomes from the two recognition techniques was compared. In this approach the outcome with a higher confidence level was chosen. If the recognition using SVM detected anger as the emotion across 10 frames with a confidence level of 0.4 and the template based recognition detected happiness as the emotion across 10 frames with a confidence level of 0.5 then happiness was chosen as the outcome. In case of recognition using SVM confidence level for individual modalities was calculated by number of times an emotion was detected divided by the number of frames. When each modality detects an emotion, it raises an emotion detected event. A function designated to handle the emotion detected event captures the event details such as modality that raised the event and the emotion detected. The fusion process determines the top emotions based on majority voting. The emotion that is detected in majority of the outcomes from individual modality is given the highest rank. Each emotion is assigned a rank but only the emotions that have a confidence level greater than the threshold T are considered to recognize the emotional state of the subject. Such event driven approach makes the multimodal recognition process asynchronous and independent of emotion recognition process at the individual modality level. With the even driven approach the multimodal affect recognition can be decoupled from availability of tracked feature data and modality level emotion recognition outcome.

## 5. RESULTS

Experiments were performed on 2 subjects enacting happiness, sadness, fear, surprise, disgust and anger using facial expression, hand gestures, body posture, and head gestures. Each individual enacted an emotion for 2 minutes. Each session was repeated 20 times. The experiments were conducted in a controlled environment. Results of classification rates using only SVM classifier are shown in Table I.

Table I. Classification using only SVM classifier for Happiness, Anger and Sadness

| Modality | Precision | Recall |
| --- | --- | --- |
| Face | 0.84, 0.77, 0.71 | 0.63, 0.54, 0.48 |
| Head | 0.76, 0.71, 0.64 | 0.36, 0.47, 0.45 |
| Body | 0.79, 0.80, 0.75 | 0.41, 0.61, 0.54 |
| Hand | 0.75, 0.79, 0.63 | 0.40, 0.43, 0.31 |
| Multimodal | 0.88, 0.83, 0.77 | 0.67, 0.68, 0.58 |

Comparison of results between individual modalities indicates that the recognition rate was highest for the body posture modality. This is different from the results obtained in previous study (Patwardhan & Knapp, 2013) where face modality showed highest classification rates. The previous study considered only position based features whereas in this research considers temporal features have been used. The body posture recognition rates improved because the body feature points tend to move more than the facial tracked points in terms of distance and number of motions. The recognition accuracy showed improvement using multimodal affect recognition compared to results obtained using individual modality. Even in case of multiple modalities, the recognition rate was better when temporal features were used in addition to position based feature. In this research we proposed using rule based emotion templates per modality. Results of augmenting SVM recognition with emotion templates are shown below.

Table II. Classification using emotion template and SVM classifier for Happiness, Anger and Sadness

| Modality | Precision | Recall |
| --- | --- | --- |
| Face | 0.81, 0.73, 0.68 | 0.65, 0.58, 0.51 |
| Head | 0.72, 0.71, 0.59 | 0.38, 0.51, 0.46 |
| Body | 0.83, 0.86, 0.79 | 0.47, 0.62, 0.57 |
| Hand | 0.71, 0.81, 0.66 | 0.51, 0.47, 0.39 |

| Multimodal | 0.89, 0.85, 0.81 | 0.72, 0.75, 0.63 |

Classification using the emotion template based recognition and SVM classifier together improved the recall and recognition rate but degraded the precision rate for face modality. The body modality showed improved rates for the combined approach as compared to using only the supervised learning approach. In case of multimodal affect recognition, combination of SVM classifier and template based recognition results were better than the results obtained from classifier based recognition alone in terms of classification, precision and recall rates.

## 6. CONCLUSIONS

In this research a multimodal affect recognition system using Kinect was developed. The results showed that rule based emotion templates can be used for augmenting the classification rates obtained from supervised learning. The rule based template methodology is easily extensible by adding more rules. Each rule can also serve as a feature vector and can be potentially used for training a classifier. The research also showed that the decision level fusion of data from multiple modalities is suitable for combining the results from the two different recognition techniques. Temporal feature data from multiple frames improved emotion classification accuracy as compared to recognition results obtained using position based feature alone. Tracking movement of feature points in addition to location based features was used to construct more rules for the emotion templates and also improved classifier and template based detection accuracy. ARUMM 1.0 was developed to predict emotions and visually display the top three emotions having empirical probability above the confidence level threshold. Each emotion was assigned a rank based on confidence level. This ability to predict multiple emotions instead of only single emotion output, from user activity and gestures would certainly be useful in making emotion aware computer systems functionally effective and appealing.

**REFERENCES**

Denis Amelynck, Maarten Grachten, Leon van Noorden and Marc Leman. 2011. Towards E-Motion Based Music Retrieval - A study of Affective Gesture Recognition. *IEEE Transactions on Affective Computing*, 99, 1.

Rafael A. Calvo and Sidney D'Mello. 2010. Affect Detection: An Interdisciplinary Review of Models, Methods and Their Applications. *IEEE Transactions on Affective Computing*, 1, 1, 18-37.

Lawrence S. Chen, Thomas S. Huang, Hai Tao, Tsutomu Miyasato and Ryohei Nakatsu. 1998. Emotion Recognition from AudioVisual Information. In *Proceedings of the IEEE Workshop on Multimedia Signal Processing*, 83-88.

Jeffrey F. Cohn, Lawrence Ian Reed, Tsuyoshi Moriyama, Jing Xiao, Karen L. Schmidt and Zara Ambadar 2004. Multimodal Coordination of Facial Action, Head Rotation, and Eye Motion during Spontaneous Smiles. In *Proceedings of the Sixth IEEE International Conference on Automatic Face and Gesture Recognition*, 129-138.

Paul Ekman. 1999, "Basic emotions," Cognition, 98, 1992, 45-60.

Hatice Gunes and Massimo Piccardi. 2005. Fusing face and body display for bi-modal emotion recognition: Single frame analysis and multi-frame post integration. *Affective Computing and Intelligent Interaction*, 3784, 102-111.

Alejandro Jaimes and Nicu Sebe. 2007. Multimodal Human Computer Interaction: A Survey. *Computer Vision and Image Understanding*, 108, 1-2, 116-134.

Ashish Kapoor and Rosalind W. Picard. 2005. Multimodal Affect Recognition in Learning Environments. In *Proceedings of the ACM International Conference on Multimedia*, 677-682.

Loic Kessous, Ginevra Castellano and George Caridakis. 2009, Multimodal Emotion Recognition in Speech-Based Interaction using Facial Interaction Using Facial Expression, Body Gesture and acoustic analysis. *Journal on Multimodal User Interfaces*, 3, 33-48.

Irene Kotsia and Ioannis Pitas. 2007. Facial Expression Recognition in Image Sequences Using Geometric Deformation Features and Support Vector Machines. *IEEE Transactions on Image Processing*, 16, 1, 172-187.

Zhi Li and Ray A. Jarvis. 2009. A multi-modal gesture recognition system in a Human-Robot Interaction scenario. *IEEE International Workshop on Robotic and Sensors Environments*, 41-46.
Amol S. Patwardhan and Gerald M. Knapp. 2013. Multimodal Affect Analysis for Product Feedback Assessment. In *Proceedings of the 2013 Industrial and Systems Engineering Research* Conference (*ISERC' 2013*), San Juan, PR, 178-186.
Rosalind W. Picard. 1997. *Affective Computing*. MIT Press, 93-94.
Ahmad Rabie, Britta Wrede, Thurid Vogt and Marc Hanheide. 2000. Evaluation and Discussion of Multi-modal Emotion Recognition. In *Proceedings of the Second International Conference on Computer and Electrical Engineering*, 1, 598- 602.
Ioanna-Ourania Stathopoulou and George A. Tsihrintzis. 2011. Emotion Recognition from Body Movements and Gestures. *Intelligent Interactive Multimedia*, 295-303.
Suprijanto, Linda Sari, Vebi Nadhira, IGN. Merthayasa and Farida I.Muchtadi. 2009. Development System for Emotion Detection Based on Brain Signals and Facial Images. *Power*, 38, 1, 320-327.
Shawna C.G Tan and Alexander Nareyek. 2009. Integrating Facial, Gesture, and Posture Emotion Expression for a 3D Virtual Agent. In *Proceedings of the 14th. International Conference on Computer Games: AI, Animation, Mobile, Interactive Multimedia, Educational & Serious Games*, 23-31.
Michel F. Valstar and Maja Pantic. 2007. How to Distinguish Posed from Spontaneous Smiles using Geometric Features. In *Proceedings of ACM International Conference on Multimodal Interfaces*. 38-45.
A. S. Patwardhan, 2016. "Structured Unit Testable Templated Code for Efficient Code Review Process", PeerJ Computer Science (in review), 2016.
A. S. Patwardhan, and R. S. Patwardhan, "XML Entity Architecture for Efficient Software Integration", International Journal for Research in Applied Science and Engineering Technology (IJRASET), vol. 4, no. 6, June 2016.
A. S. Patwardhan and G. M. Knapp, "Affect Intensity Estimation Using Multiple Modalities," Florida Artificial Intelligence Research Society Conference, May. 2014.
A. S. Patwardhan, R. S. Patwardhan, and S. S. Vartak, "Self-Contained Cross-Cutting Pipeline Software Architecture," International Research Journal of Engineering and Technology (IRJET), vol. 3, no. 5, May. 2016.
A. S. Patwardhan, "An Architecture for Adaptive Real Time Communication with Embedded Devices," LSU, 2006.
A. S. Patwardhan and G. M. Knapp, "Multimodal Affect Analysis for Product Feedback Assessment," IIE Annual Conference. Proceedings. Institute of Industrial Engineers-Publisher, 2013.
A. S. Patwardhan and G. M. Knapp, "Aggressive Action and Anger Detection from Multiple Modalities using Kinect", submitted to ACM Transactions on Intelligent Systems and Technology (ACM TIST) (in review).
A. S. Patwardhan and G. M. Knapp, "EmoFit: Affect Monitoring System for Sedentary Jobs," preprint, arXiv.org, 2016.
A. S. Patwardhan, J. Kidd, T. Urena and A. Rajagopalan, "Embracing Agile methodology during DevOps Developer Internship Program", IEEE Software (in review), 2016.
A. S. Patwardhan, "Analysis of Software Delivery Process Shortcomings and Architectural Pitfalls", PeerJ Computer Science (in review), 2016.